\documentclass[%
 reprint,
 amsmath,amssymb,
 aps,
]{revtex4-2}

\usepackage{graphicx}
\usepackage{dcolumn}
\usepackage{bm}
\usepackage{lineno}
\usepackage{subfigure}

\begin{document}

\preprint{APS/123-QED}

\title{Precise Measurements of Decay Parameters and $CP$ Asymmetry\\ with Entangled $\Lambda-\bar{\Lambda}$ pairs} 

\author{\small 
M.~Ablikim$^{1}$, M.~N.~Achasov$^{11,b}$, P.~Adlarson$^{70}$, M.~Albrecht$^{4}$, R.~Aliberti$^{31}$, A.~Amoroso$^{69A,69C}$, M.~R.~An$^{35}$, Q.~An$^{66,53}$, X.~H.~Bai$^{61}$, Y.~Bai$^{52}$
% ...
\\
\vspace{0.2cm}
(BESIII Collaboration)\\
\vspace{0.2cm} {\it
$^{1}$ Institute of High Energy Physics, Beijing 100049, People's Republic of China\\
% ...
$^{4}$ Bochum Ruhr-University, D-44780 Bochum, Germany\\
% ...
$^{11}$ G.I. Budker Institute of Nuclear Physics SB RAS (BINP), Novosibirsk 630090, Russia\\
% ...
$^{31}$ Johannes Gutenberg University of Mainz, Johann-Joachim-Becher-Weg 45, D-55099 Mainz, Germany\\
% ...
$^{35}$ Liaoning Normal University, Dalian 116029, People's Republic of China\\
% ...
$^{52}$ Southeast University, Nanjing 211100, People's Republic of China\\
$^{53}$ State Key Laboratory of Particle Detection and Electronics, Beijing 100049, Hefei 230026, People's Republic of China\\
% ...
$^{61}$ University of Jinan, Jinan 250022, People's Republic of China\\
% ...
$^{66}$ University of Science and Technology of China, Hefei 230026, People's Republic of China\\
% ...
$^{69}$ University of Turin and INFN, (A)University of Turin, I-10125, Turin, Italy; (B)University of Eastern Piedmont, I-15121, Alessandria, Italy; (C)INFN, I-10125, Turin, Italy\\
$^{70}$ Uppsala University, Box 516, SE-75120 Uppsala, Sweden\\
% ...
$^{b}$ Also at the Novosibirsk State University, Novosibirsk, 630090, Russia\\
% ...
}}

\vspace{0.4cm}

\date{\today}

\begin{abstract}
Based on 10 billion $J/\psi$ events collected at the BESIII experiment,  
a search for $CP$ violation in $\Lambda$ decay is performed in the difference between $CP$-odd decay parameters $\alpha_{-}$ for $\Lambda \rightarrow p\pi^-$
and $\alpha_{+}$ for $\bar\Lambda \rightarrow \bar{p}\pi^+$ by using the  process $e^+e^- \to J/\psi \rightarrow \Lambda \bar\Lambda $.
With a five-dimensional fit to the full angular distributions of the daughter baryon,  
the most precise values for the decay parameters are determined to be $\alpha_{-} = 0.7519 \pm 0.0036 \pm 0.0024$
and $\alpha_{+} = -0.7559 \pm 0.0036 \pm 0.0030$, respectively.  The $\Lambda$ and $\bar{\Lambda}$ averaged value of the decay parameter is extracted 
 to be $\alpha_{\rm{avg}} = 0.7542 \pm 0.0010 \pm 0.0024$ with unprecedented accuracy.   
The $CP$ asymmetry  $A_{CP}=(\alpha_{-}+\alpha_{+})/(\alpha_{-}-\alpha_{+})$
is determined to be $-0.0025 \pm 0.0046 \pm 0.0012$, which is one of the most precise measurements 
in the baryon sector.  The reported results for the decay parameter will play an important role in the studies of the polarizations and $CP$ violations for the strange, charmed and beauty baryons.   
\end{abstract}

\maketitle


Charge-parity ($CP$) violation is a subject of continuing interest. 
To date, $CP$ violation has been discovered in the $K$~\cite{PhysRevLett.13.138}, 
$B$~\cite{BaBar:2001pki,Belle:2001zzw}, and $D$ meson~\cite{LHCb:2019hro} systems, 
but it has never been observed in the decays of any baryon. 
Hence it is vital to search for additional  sources of $CP$ violation. 
Moreover, the Standard Model(SM) is difficult to explain the phenomenon of matter-antimatter asymmetry 
of the universe~\cite{Sakharov:1967dj, Bernreuther:2002uj, Canetti:2012zc}. 
Thus, the $CP$ test is an ideal and sensitive method to search for the physics beyond SM~\cite{Bigi:2000yz, Bediaga:2020qxg}. 


The promising $CP$-violating signature in spin-$\frac{1}{2}$ nonleptonic hyperon decays
is the difference between hyperon and antihyperon decay distributions in their parity-violating two-body weak
decays~\cite{Lee:1957qs}. In such decays the angular distribution of the
daughter baryon is proportional to $\left(1+\alpha_{Y} \bf{P}_{Y} \cdot \hat{p}_{d}\right)$, where $\alpha_{Y}$ is 
the hyperon decay parameter, and $\bf{P}_{Y}$ and $\bf{\hat p}_{d}$ are the hyperon polarization and the unit vector in the direction of the daughter baryon momentum, respectively, both in the hyperon rest frame. The $CP$ asymmetry is defined as 
$A_{CP}=\frac{\alpha_{Y}+\alpha_{\bar{Y}}}{\alpha_{Y}-\alpha_{\bar{Y}}}$. 
The parameters $\alpha_{Y}$ and $\alpha_{\bar{Y}}$ are $CP$ odd so that
a nonzero $A_{CP}$ indicates $CP$ violation. 
In the SM, the Cabibbo-Kobayashi-Maskawa mechanism predicts a tiny $A_{CP}$ value of 
$\sim\! 10^{-4}$~\cite{Donoghue:1985ww, *PhysRevD.34.833}.  
Therefore, the hyperon decay 
is sensitive to the sources of $CP$ asymmetry from physics beyond the SM~\cite{BESIII:2018cnd, Ireland:2019uja}. 
A precise measurement of the $\Lambda(\bar{\Lambda})$ decay parameters is important for studies of 
spin polarization\cite{STAR:2019erd, Becattini:2020ngo, BESIII:2019nep, STAR:2021beb} 
and decay parameters\cite{Blake:2019guk, Han:2019axh, BESIII:2019odb, FOCUS:2005vxq, LHCb:2020iux, Wang:2016elx, BESIII:2020kap} 
of many other baryons($\Sigma^{0}, \Xi^{0}, \Xi^{-}, \Omega^{-}, \Lambda_{c}, \Lambda_{b}$ etc.) decays into final 
states involving $\Lambda$.  

 A total sample of 10 billion $J/\psi$ events has been collected by the BESIII experiment,  
 about 3.2 million quantum-entangled $\Lambda$-$\bar{\Lambda}$ pairs are expected to be fully 
 reconstructed in the decay $J/\psi \rightarrow \Lambda \bar\Lambda $ with $\Lambda (\bar\Lambda ) 
 \rightarrow p\pi^- (\bar{p}\pi^+)$~\cite{Li:2016tlt}. Hence, in this Letter,  we present the most precise measurements 
 of $\Lambda$ decay parameters and  $CP$ asymmetry with a five-dimensional fit to the full 
 angular distributions of the daughter baryon.  
  
Two real parameters,  the $J/\psi \rightarrow \Lambda \bar{\Lambda}$ angular distribution parameter $\alpha_{J/\psi}$ 
and the helicity phase difference $\Delta\Phi$, describe the angular distribution and polarization of the produced $\Lambda$ and $\bar{\Lambda}$~\cite{Dubnickova:1992ii,*Gakh:2005hh,*Czyz:2007wi,*Faldt:2013gka,*Faldt:2016qee,*Faldt:2017kgy}. 
If the phase difference $\Delta \Phi$ is nonvanishing, 
the polarization of the $\Lambda$ and $\bar{\Lambda}$ 
will be oriented perpendicular to the production plane. 
For the decay $\Lambda \rightarrow p \pi^{-}$, the angular distribution of the proton is 
$\frac{1}{4 \pi}\left(1+\alpha_{-} \mathbf{P}_{\Lambda} \cdot \hat{\mathbf{n}}\right)$, 
where  $\alpha_{-}$ is $\Lambda$ decay parameter, $\mathbf{P}_{\Lambda}$ is the polarization vector of the $\Lambda$, 
$\hat{\mathbf{n}}$ is the unit vector of the proton momentum in the $\Lambda$ rest frame. The definition of the decay parameter 
$\alpha_{+}$ for $\bar{\Lambda} \rightarrow \bar{p} \pi^{+}$ follows an analogous convention~\cite{Workman:2022ynf}. 

\begin{figure}[!htbp]
    \centering 
    \includegraphics[width=1.0\linewidth]{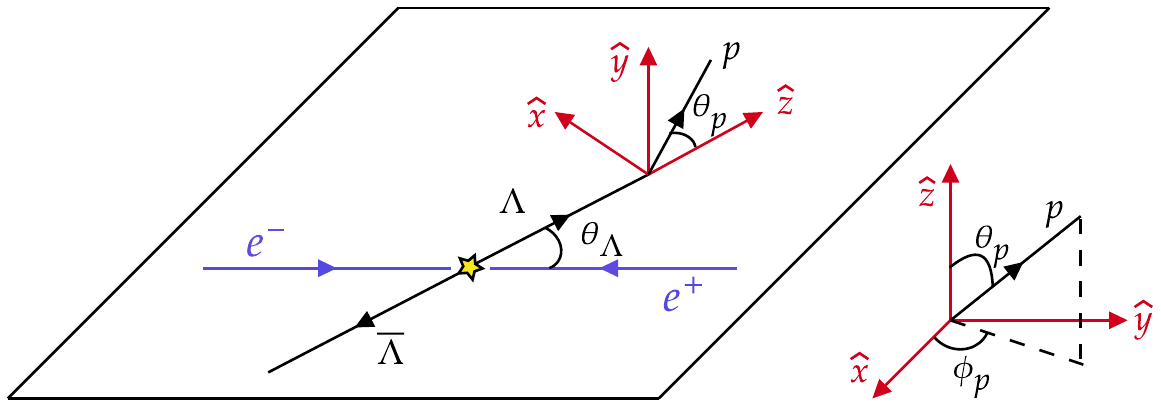}
    \caption{ Definition of the right-hand coordinate system used to describe the $J/\psi \rightarrow \Lambda\bar{\Lambda}$ process. 
             In the $e^+e^-$ center-of-mass system, the $\Lambda$ is emitted along the $\hat{z}$ axis direction. 
             $\hat{y}$ axis is perpendicular to the plane of $\Lambda$ and $e^{-}$.
             The hyperons are polarized along the $\hat{y}$ direction.} 
    \label{fig:diagram_of_decay}
\end{figure}

For the cascade decay $J/\psi \rightarrow \Lambda \bar\Lambda $ with $\Lambda (\bar\Lambda ) \rightarrow p\pi^- (\bar{p}\pi^+)$, 
the angular distribution of each event is uniquely characterized by the kinematic variable
$\xi = (\theta_{\Lambda}, \theta_{p}, \phi_{p}, \theta_{\bar{p}}, \phi_{\bar{p}})$, 
where $\theta_{p}, \phi_{p}$ and $\theta_{\bar{p}}, \phi_{\bar{p}}$ 
are the polar and azimuth angles of the proton and antiproton 
in their mother particles rest frames. The components of these vectors are expressed using a 
right-handed coordinate system 
$(\hat{x}, \hat{y}, \hat{z})$ shown in Fig.~\ref{fig:diagram_of_decay}.
The $\hat{z}$ axis is taken along the $\Lambda$ momentum 
$\mathbf{p}_{\Lambda} =-\mathbf{p}_{\bar{\Lambda}}\equiv\mathbf{p}$ 
in the $e^{+}e^{-}$ center-of-mass system (CMS). 
The $\hat{y}$ axis is perpendicular to the production plane and oriented along the vector 
$\mathbf{k} \times \mathbf{p}$, where $\mathbf{k}_{e^{-}} =-\mathbf{k}_{e^{+}}\equiv\mathbf{k}$ 
is the electron beam momentum. 
The scattering angle of the $\Lambda$ is given by 
$\cos \theta_{\Lambda}=\hat{\mathbf{p}} \cdot \hat{\mathbf{k}}$.

The differential distribution function can be expressed as 
 \begin{equation}
     \begin{aligned} 
         \mathcal{W}&(\xi) = \mathcal{F}_{0}(\xi) + \alpha_{J/\psi} \mathcal{F}_{5}(\xi) + \alpha_{-} \alpha_{+} \\ 
         & \times \left[\mathcal{F}_{1}(\xi) + \sqrt{1-\alpha_{J/\psi}^{2}} \cos(\Delta \Phi) 
         \mathcal{F}_{2}(\xi) + \alpha_{J/\psi} \mathcal{F}_{6}(\xi) \right] \\ 
         & + \sqrt{1-\alpha_{J/\psi}^{2}} \sin(\Delta \Phi) \left[\alpha_{-} \mathcal{F}_{3}(\xi) + \alpha_{+} 
         \mathcal{F}_{4}(\xi) \right] \\ 
     \end{aligned}
     \label{eq:amplitude}
 \end{equation}
where the angular functions $\mathcal{F}_{i}(\xi)\ (i=0,1,...6)$ are described in detail in Ref.~\cite{Faldt:2017kgy}.
The terms proportional to $\alpha_{-} \alpha_{+}$ in Eq.~\eqref{eq:amplitude} represent the contribution from $\Lambda$--$\bar{\Lambda}$ 
spin correlations, and the terms proportional to $\alpha_{-}$ and $\alpha_{+}$ separately represent the contribution from the hyperon 
transverse polarization $P_{y}$, defined as:
\begin{equation}
    P_{y}\left(\cos \theta_{\Lambda}\right)=\frac{\sqrt{1-\alpha_{J/\psi}^{2}} 
    \sin (\Delta \Phi) \cos \theta_{\Lambda} \sin \theta_{\Lambda}}{1+\alpha_{J/\psi} 
    \cos ^{2} \theta_{\Lambda}}.
\end{equation}

The analysis presented here is based on the aforementioned sample of $10$~billion $J/\psi$ events~\cite{BESIII:2021cxx} 
collected at the BESIII detector~\cite{BESIII:2009fln, BESIII:2020nme}. A Monte Carlo (MC) simulation of  
$J/\psi$ samples is used to determine the detector efficiency, optimize the event selection,  
and estimate the background. The simulation is performed by the GEANT4-based~\cite{GEANT4:2002zbu} BESIII Object 
Oriented Simulation Tool project~\cite{Deng:2006}, 
which includes the geometric description of the BESIII detector and the detector response. 
The MC event generators KKMC~\cite{Jadach:2000ir}, BesEvtGen~\cite{Ping:2008zz}, 
and Lundcharm~\cite{Chen:2000tv,Yang:2014vra} are used to describe $J/\psi$ production together with known and 
unknown decay modes. For the signal process, 
$J/\psi \rightarrow \Lambda \bar{\Lambda}$, the parameters of the angular distribution are 
obtained from previous measurements~\cite{BESIII:2018cnd}. For the dominant background channel 
$J/\psi \rightarrow \gamma \eta_{c}(\eta_{c} \rightarrow \Lambda \bar{\Lambda})$, 
the decay $J/\psi \rightarrow \gamma \eta_{c}$ is generated with an angular distribution 
of $1+\rm{cos}^{2}\theta_{\gamma}$~\cite{Liao:2009zz}, 
where $\theta_{\gamma}$ is the angle between the photon and positron beam direction 
in the CMS, and another background channel  
$J/\psi \rightarrow \gamma \Lambda \bar{\Lambda}$ is described by the phase space model.

The $\Lambda$ and $\bar{\Lambda}$ baryons are reconstructed from their dominant hadronic decay mode, 
$\Lambda(\bar{\Lambda}) \rightarrow p\pi^{-}(\bar{p}\pi^{+})$. Charged tracks detected in the main drift chamber must satisfy 
$\lvert \cos(\theta) \rvert < 0.93$ , where $\theta$ is the angle 
between the charged track and the positron beam direction. 
Events with at least four charged tracks are retained. 
Tracks with momentum larger than 
$0.5\ \text{GeV}/c$ are considered as proton candidates, otherwise as pion candidates.
There are no further particle identification requirements.
Vertex fits are performed by looping over 
all combinations with oppositely charged proton and pion candidates, constraining them to a 
common vertex. The pairs with vertex fit $\chi^{2}$ 
lower than $200$ and decay length larger than $0$
are regarded as $\Lambda-\bar{\Lambda}$ candidates. 
A four-momentum constrained kinematic fit is applied to the
$p\bar{p}\pi^+\pi^-$ hypothesis, and events with a minimum $\chi^2$ lower than $60$ are 
selected as $J/\psi$ candidates. 

\begin{figure}[!htbp]
    \includegraphics[width=0.85\linewidth]{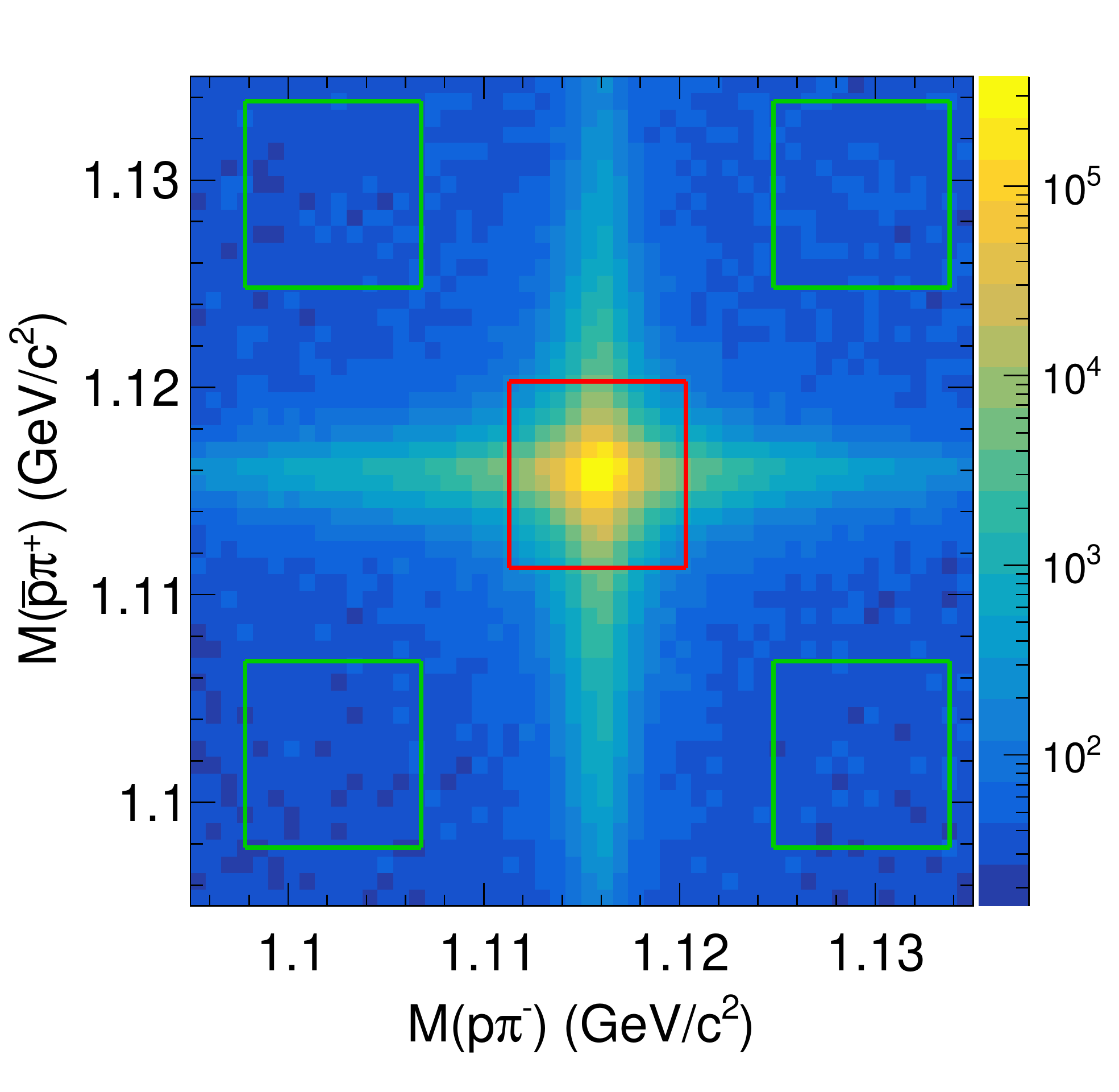}
 \caption{\label{fig:lbd_mass}%
    Distribution of the invariant mass spectra of $\bar{p}\pi^{+}$ versus the invariant mass 
    spectra of $p\pi^{-}$ from the data. The signal and sideband regions are denoted by 
    red and green boxes, respectively. The z axis is in logarithm style.
 }%
\end{figure}

An inclusive MC sample of $10$ billion $J/\psi$ events is used for studying potential backgrounds. After applying the same selection criteria as for the data,
the main backgrounds are divided into two types according to the shapes of $m_{p\pi^{-}}$ and $m_{\bar{p}\pi^{+}}$: 
(1)BKGI, nonpeaking backgrounds, including $J/\psi \rightarrow p\pi^-\bar{p}\pi^+$, 
$\Delta^{++}\bar{p}\pi^-$, $\bar{\Delta}^{++}p\pi^+$, $\Delta^{++}\bar{\Delta}^{++}$;
(2)BKGII, peaking backgrounds, 
including $J/\psi \rightarrow \gamma \Lambda \bar{\Lambda}$, $\gamma \eta_{c}
(\eta_{c} \rightarrow \Lambda \bar{\Lambda})$. 
The number of nonpeaking backgrounds is estimated by the two-dimensional sideband 
regions of the $m_{p\pi^{-}}$ versus $m_{\bar{p}\pi^{+}}$ distribution from the data sample 
which is shown in Fig.~\ref{fig:lbd_mass}.
The signal region is defined as $m_{p\pi^{-}/\bar{p}\pi^{+}} \in [1.111, 1.121]\ \text{GeV}/c^2$, 
and the lower and higher sideband regions are defined as $m_{p\pi^{-}/\bar{p}\pi^{+}} \in [1.098, 1.107]$ GeV/$c^2$ 
and $m_{p\pi^{-}/\bar{p}\pi^{+}} \in [1.125, 1.134]$ GeV/$c^2$, respectively. 
The yields of various peaking background sources 
are estimated by individual exclusive MC samples, then normalized to the data sample 
according to their branching fractions~\cite{Workman:2022ynf}.
The final data sample contains $3231781$ events including the estimated  
background yield of $3801 \pm 63$ events. 
The sample has a high purity of $99.9\%$.

Based on the joint angular distribution, a maximum likelihood fit with four free parameters ($\alpha_{J/\psi}$, $\Delta\Phi$, $\alpha_{-}$, and $\alpha_{+}$) is performed. The joint likelihood function is defined as:
\begin{eqnarray}
    \mathcal{L}\ && = \prod^{N}_{i=1}\mathcal{P}(\xi^i;\alpha_{J/\psi},\Delta\Phi,\alpha_-,\alpha_+) \nonumber \\
                 && = \prod^{N}_{i=1}\mathcal{C}\mathcal{W}(\xi^i;\alpha_{J/\psi},
                \Delta\Phi,\alpha_-,\alpha_+)\epsilon(\xi^i)~,
    \label{eq:ML_likelihood}
\end{eqnarray}
where $\mathcal{P}(\xi^i;\alpha_{J/\psi},\Delta\Phi,\alpha_-,\alpha_+)$ is 
the probability density function of $\xi^i$, 
the kinematic variable of event $i$, and $\mathcal{W}(\xi^i;\alpha_{J/\psi},\Delta\Phi,\alpha_-,\alpha_+)$
is given by Eq.~\eqref{eq:amplitude}.
The detection efficiency is denoted by $\epsilon(\xi^i)$. 
The normalization factor $\mathcal{C}^{-1} = \frac{1}{N_{\rm{MC}}}\sum^{N_{\rm{MC}}}_{j=1} 
\mathcal{W}(\xi^j;\alpha_{J/\psi},\Delta\Phi,\alpha_-,\alpha_+)\epsilon(\xi^j)$
is estimated with the $N_{\rm{MC}}$ events generated with the phase space model, 
applying the same event selection criteria as for the data.
To improve the accuracy of the normalization factor, we generate a MC sample about 100 times larger than the selected experimental data.
We use the ROOFIT package~\cite{Verkerke:2003ir} to determine the fit
parameters from the minimization of the function:
\begin{eqnarray}
    \mathcal{S} = -\rm{ln}\mathcal{L}_{data} + \rm{ln}\mathcal{L}_{BKGI} + \rm{ln}\mathcal{L}_{BKGII}~,
\end{eqnarray}
where $\mathcal{L}_{\rm{data}}$ and $\mathcal{L}_{\rm{BKGI}}$ are the likelihood function of events in the signal region 
and sideband regions, respectively. The $\mathcal{L}_{\rm{BKGII}}$ is the likelihood function of background events 
obtained by exclusive MC samples. 
The likelihood function of background event is the same as the data.
The results of the maximum likelihood fit of data are given in Table~\ref{tab:final_results}, 
with the $CP$ asymmetry  given by $A_{CP} = (\alpha_{-}+\alpha_{+})/(\alpha_{-}-\alpha_{+})$, 
and the average value of the $\Lambda$ and $\bar{\Lambda}$ 
decay parameters $\alpha_{\rm{avg}} = (\alpha_{-}-\alpha_{+})/2$.  
The correlation coefficient between $\alpha_{-}$ and $\alpha_{+}$ is $\rho(\alpha_{-}, \alpha_{+}) = 0.850$. 
\begin{table}[h]
\caption{
\label{tab:final_results}
	The angular distribution parameters, $\alpha_{J/\psi}$, $\Delta\Phi$
	and the asymmetry parameters $\alpha_-$ for $\Lambda \rightarrow p\pi^-$, $\alpha_+$
	for $\bar{\Lambda} \rightarrow \bar{p}\pi^+$ obtained in this work and in
    previous BESIII measurements~\cite{BESIII:2018cnd} for comparison.
    The first uncertainty is statistical, the second one is systematic.
    }
\begin{ruledtabular}
\begin{tabular}{ccc}
    \multicolumn{1}{r}{\textrm{Par.}}&
    \multicolumn{1}{c}{\textrm{This work}}&
    \multicolumn{1}{c}{\textrm{Previous results~\cite{BESIII:2018cnd}}}\\
\colrule
$\alpha_{J/\psi}$ &  0.4748 $\pm$ 0.0022 $\pm$ 0.0031 & 0.461 $\pm$ 0.006 $\pm$ 0.007\\
$\Delta\Phi$      &  0.7521 $\pm$ 0.0042 $\pm$ 0.0066 & 0.740 $\pm$ 0.010 $\pm$ 0.009\\
$\alpha_{-}$      &  0.7519 $\pm$ 0.0036 $\pm$ 0.0024 & 0.750 $\pm$ 0.009 $\pm$ 0.004\\
$\alpha_{+}$      &  $-$0.7559 $\pm$ 0.0036 $\pm$ 0.0030 & $-$0.758 $\pm$ 0.010 $\pm$ 0.007\\
\hline
$A_{CP}$          & $-$0.0025 $\pm$ 0.0046 $\pm$ 0.0012 & 0.006 $\pm$ 0.012 $\pm$ 0.007\\
    $\alpha_{\rm{avg}}$    &  0.7542 $\pm$ 0.0010 $\pm$ 0.0024 &  - \\
\end{tabular}
\end{ruledtabular}
\end{table}

\begin{figure}[h]
    \includegraphics[width=0.9\linewidth]{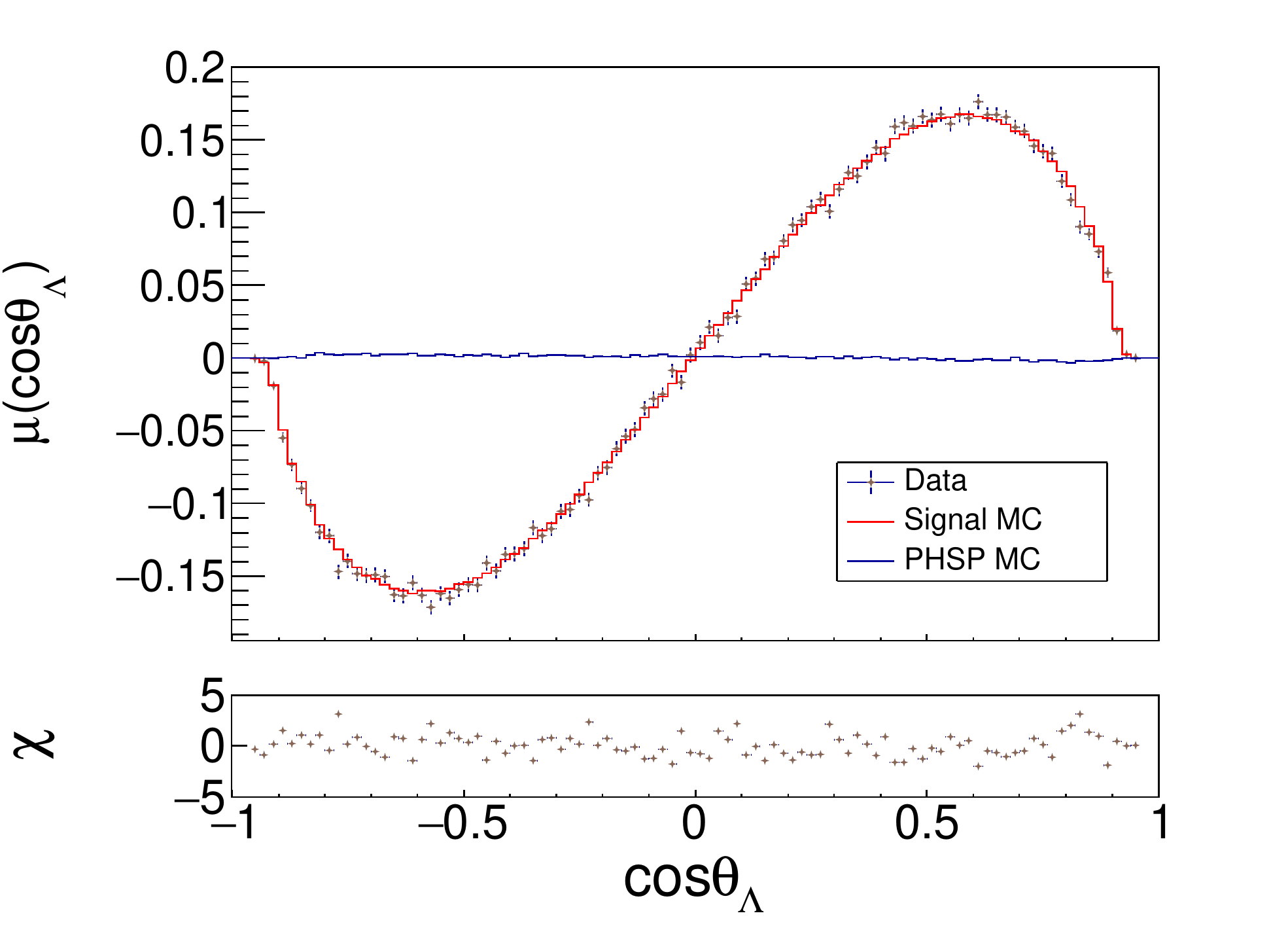}
    \caption{\label{fig:mu_distribution} 
    Distribution of moment $\mu(\cos \theta_{\Lambda})$ versus 
	$\cos \theta_{\Lambda}$.
	The points with error bars are
    the data, and the red histogram is the signal MC sample with input parameters fixed to fit results. 
    The blue histogram shows the result from phase space(PHSP) MC sample. 
    The distribution of $\chi = (\mu_{\rm{data}}-\mu_{\rm{MC}})/\sigma(\mu_{\rm{data}})$ is shown at the bottom, 
    where $\mu_{\rm{data}}$ and $\mu_{\rm{MC}}$ are the moments of data and signal MC sample. The 
    $\sigma(\mu_{\rm{data}})$ is the statistical uncertainty of $\mu_{\rm{data}}$.}
\end{figure}

The moment
\begin{eqnarray}
    \mu[\cos(\theta_{\Lambda})]=(m/N) \sum^{\emph{N}_{k}}_{i=1}(n_{1, y}^{(i)}-n_{2, y}^{(i)})~,
\end{eqnarray}
which related to the polarization, is used to compare the consistency 
between the data and the fit results.
Hereby, $N$ is the total number of events in the data set, and $m = 100$ is the number of bins in $\cos(\theta_{\Lambda})$ for calculating the moment. 
$N_{k}$ denotes the number of events in the $k$ th $\cos(\theta_{\Lambda})$ bin.
The expected angular dependence of the moment for the acceptance-corrected data reads
\begin{eqnarray}
\label{eq:mu}
    \mu\left(\cos \theta_{\Lambda}\right)=\frac{\alpha_{-}-\alpha_{+}}{2} 
    \frac{1+\alpha_{J/\psi} \cos ^{2} \theta_{\Lambda}}{3+\alpha_{J/\psi}} P_{y}\left(\theta_{\Lambda}\right)~.
\end{eqnarray}
A significant transverse polarization of $\Lambda$ and $\bar{\Lambda}$ can 
be seen in Fig.~\ref{fig:mu_distribution}, in which the points with error bars are the data, and 
the solid line is obtained from signal MC sample generated by Eq.~\eqref{eq:amplitude}, 
where the input parameters are taken from fit results.
The data are consistent with the fit results. 

The systematic uncertainties in this analysis can be divided into two categories: 
(A) the uncertainties from event selection, including background estimation, tracking, the $\Lambda/\bar{\Lambda}$ vertex fit and kinematic fit; 
(B) the uncertainty associated with the fit procedure.
The uncertainty from background is estimated by varying 
the input background numbers by 1 standard deviation. 
The differences on the fitted parameters are taken as the systematic uncertainty. 
For the tracking and $\Lambda-\bar{\Lambda}$ vertex fit and kinematic fit, 
a correction to the MC efficiency is made. 
We use control samples to get the efficiencies of the data and the MC simulation 
in tracking, $\Lambda-\bar{\Lambda}$ vertex fit, and kinematic fit, 
and use the data and MC difference to calibrate the MC sample. 
The uncertainty due to the charged particle tracking efficiency has been investigated 
with a $J/\psi \rightarrow p\pi^{-}\bar{p}\pi^{+}$ 
control sample. The systematic uncertainties due to the $\Lambda$ and $\bar{\Lambda}$ 
vertex reconstruction and kinematic fits are estimated by a control sample 
$J/\psi \rightarrow \Lambda \bar{\Lambda} \rightarrow p\pi^{-}\bar{p}\pi^{+}$.
In order to reduce the impact of statistical fluctuations, the fit with a corrected MC sample
is performed $100$ times by varying the correction factor randomly within 1 standard deviation.
The differences between the mean value of the fit results with corrections and the nominal fit are taken as the systematic uncertainties.
The MC simulation is used to estimate the uncertainty of the fit method. 
The sum of the differences between the input and output values and their uncertainty 
are regarded as systematic uncertainties.
The absolute systematic uncertainties for various sources are summarized in Table~\ref{tab:sys_sum}. 
The total systematic uncertainty of each 
parameter is obtained by summing the individual contributions in quadrature.

\begin{table}[h]
\caption{
\label{tab:sys_sum}
	Absolute systematic uncertainties for the measured parameters $\alpha_{J/\psi}$, $\Delta\Phi$, 
	the asymmetry parameters $\alpha_-$ for $\Lambda \rightarrow p\pi^-$, $\alpha_+$
	for $\bar{\Lambda} \rightarrow \bar{p}\pi^+$, 
    the $CP$ asymmetry value $A_{CP}$ 
    and the average value of the $\Lambda$ asymmetry parameter $\alpha_{\rm{avg}}$.  
    }
\begin{ruledtabular}
\begin{tabular}{ccccccc}
    \multicolumn{1}{c}{\textrm{Source ($10^{-3}$)}}&
    \multicolumn{1}{c}{\textrm{$\alpha_{J/\psi}$}}&
    \multicolumn{1}{c}{\textrm{$\Delta\Phi$}}&
    \multicolumn{1}{c}{\textrm{$\alpha_{-}$}}&
    \multicolumn{1}{c}{\textrm{$\alpha_{+}$}}&
    \multicolumn{1}{c}{\textrm{$A_{CP}$}}&
    \multicolumn{1}{c}{\textrm{$\alpha_{\rm{avg}}$}}\\
\colrule
     Background 
     & 0.4 & 0.2 & 0.3 & 0.4 & 0.4 & 0.0 \\
     Tracking 
     & 1.7 & 0.6 & 1.7 & 2.1 & 0.2 & 1.9 \\
     $\Lambda/\bar{\Lambda}$ vertex fit 
     & 0.2 & 0.0 & 0.1 & 0.3 & 0.3 & 0.1 \\
     Kinematic fit      
     & 1.4 & 3.0 & 0.8 & 1.4 & 0.3 & 0.4 \\
     Fit method      
     & 2.1 & 5.8 & 1.5 & 1.6 & 1.0 & 1.4 \\
     \hline
     Total          
     & 3.1 & 6.6 & 2.4 & 3.0 & 1.2 & 2.4 \\
\end{tabular}
\end{ruledtabular}
\end{table}

In summary, by analyzing $10$ billion $J/\psi$ events, 
we report the most precise measurements of the decay parameters of 
$\Lambda(\bar{\Lambda})$, with results given in Table~\ref{tab:final_results}. 
The results are consistent with those in the previous analysis~\cite{BESIII:2018cnd}, however, with 
significantly improved accuracy. 
The measured $CP$ asymmetry provides a hunting ground for physics beyond the standard model~\cite{Tandean:2003fr}. 
A clear transverse polarization is observed for the $\Lambda$ and $\bar{\Lambda}$ as shown in Fig.~\ref{fig:mu_distribution}. 
The phase between helicity flip and helicity conserving transitions is determined to be 
$\Delta\Phi=0.7521 \pm 0.0042 \pm 0.0066$, where the first uncertainty is statistical 
and the second one is systematic. 
\begin{figure}[!ht]
    \includegraphics[width=0.9\linewidth]{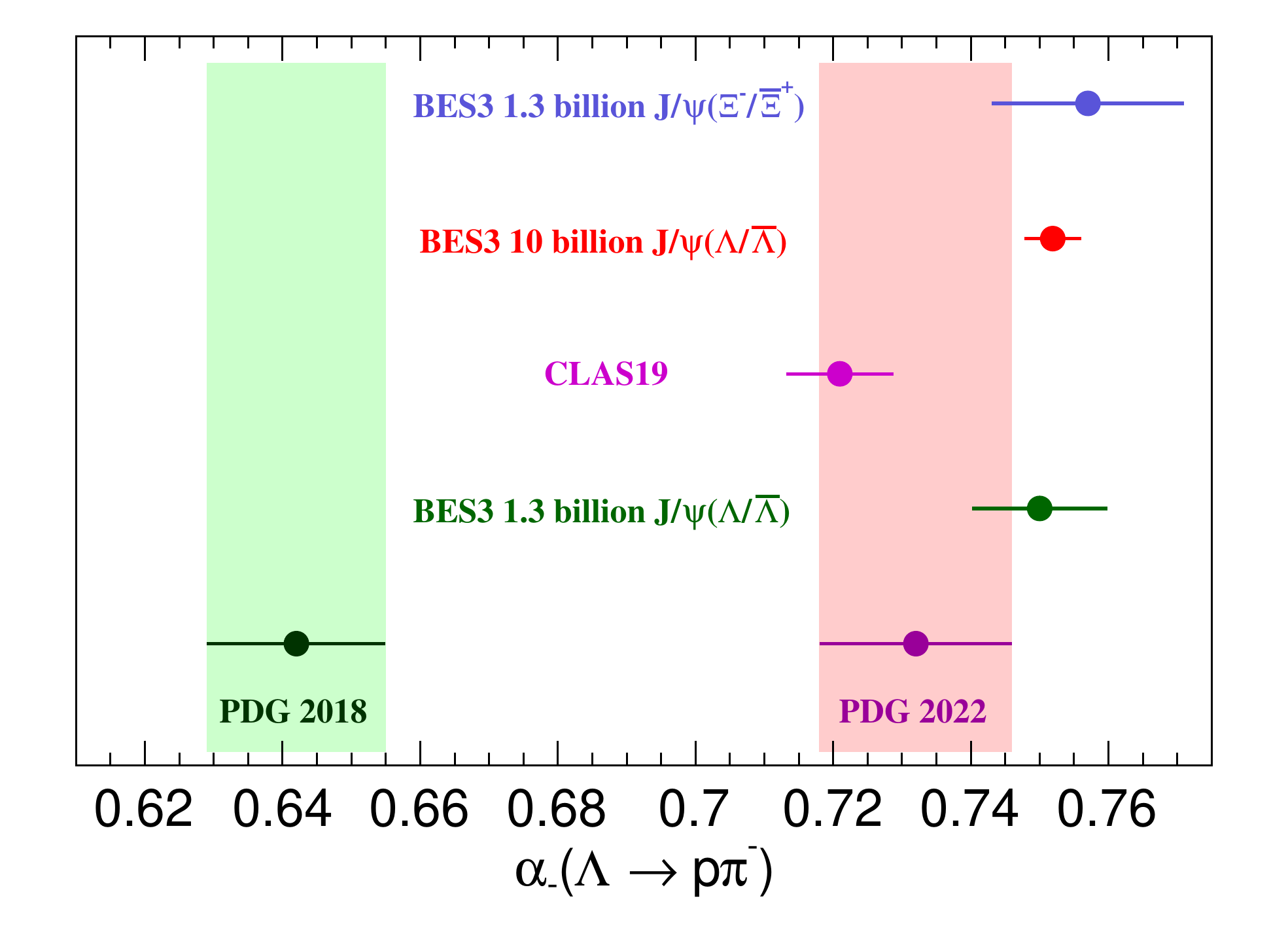} 
    \caption{\label{fig:PDG_value}
    Results of the $\Lambda$ decay parameter from different experiments. 
    The green band represents the PDG 2018 value, 
    and the pink band represents the PDG 2022 value.}
\end{figure}
The large value of the phase makes it possible to 
simultaneously determine the decay parameters of $\Lambda \rightarrow p\pi^{-}$ and $\bar{\Lambda} \rightarrow \bar{p}\pi^{+}$ to be 
$\alpha_{-} = 0.7519 \pm 0.0036 \pm 0.0024$ and 
$\alpha_{+} = 0.7559 \pm 0.0036 \pm 0.0030$, which represents the most precise 
measurements to date. 
Owing to the large correlation coefficient
of the two decay parameters $\rho(\alpha_{-}, \alpha_{+}) = 0.850$, 
the $\Lambda$ and $\bar{\Lambda}$ averaged value is determined to be 
$\alpha_{\rm{avg}} = 0.7542 \pm 0.0010 \pm 0.0024$ for the first time, which are the most precise measurements in the baryon sector.  
Being the lightest baryon with strangeness, 
the measurements of polarizations, decay parameters, and $CP$ asymmetries of heavier baryons,  
therefore, implicitly depend on $\alpha_{\Lambda}$~\cite{BESIII:2019odb, FOCUS:2005vxq, LHCb:2020iux, Wang:2016elx, BESIII:2020kap}. 

Results of the $\Lambda$ decay parameter from different experiments are shown in Fig.~\ref{fig:PDG_value}. 
The $\alpha_{-}$ value obtained in this work agrees with 
the previous BESIII measurements~\cite{BESIII:2018cnd} and the BESIII result 
extracted from the $J/\psi \rightarrow \Xi^{-}\bar{\Xi}^{+}$ decay~\cite{BESIII:2021ypr}, 
but deviates from the CLAS result by $3.5 \sigma$. 
In addition, we obtain the value of $CP$ violation for the $\Lambda$ decay   
$A_{CP}=(\alpha_{-}+\alpha_{+})/(\alpha_{-}-\alpha_{+})=-0.0025\pm0.0046\pm0.0012$, 
which is compatible with zero, thereby, indicating a non-$CP$-violation scenario.
The next generation of charm factories ~\cite{Charm-TauFactory:2013cnj, Luo:2018njj} will greatly 
improve the accuracy of the $CP$-violating measurements, and shed light on the mechanism of $CP$ violation in the baryon sector. 

\begin{acknowledgments}
The BESIII collaboration thanks the staff of BEPCII and the IHEP computing center for their strong support. 
    This work is supported in part by the National Key R$\&$D Program of China under Contracts Nos. 2020YFA0406300, 
2020YFA0406400; National Natural Science Foundation of China (NSFC) under Contracts Nos. 11635010, 11735014, 
11835012, 11935015, 11935016, 11935018, 11961141012, 12022510, 12025502, 12035009, 12035013, 12192260, 12192261, 
12192262, 12192263, 12192264, 12192265; the Chinese Academy of Sciences (CAS) Large-Scale Scientific Facility Program; 
Joint Large-Scale Scientific Facility Funds of the NSFC and CAS under Contract No. U1832207; CAS Key Research Program 
of Frontier Sciences under Contract No. QYZDJ-SSW-SLH040; 100 Talents Program of CAS; INPAC and Shanghai Key Laboratory 
for Particle Physics and Cosmology; Polish National Science Centre under Contract 2019/35/O/ST2/02907; 
ERC under Contract No. 758462; European Union's Horizon 2020 research and innovation 
programme under Marie Sklodowska-Curie grant agreement under Contract No. 894790; German Research Foundation DFG under 
Contracts Nos. 443159800, Collaborative Research Center CRC 1044, GRK 2149; Istituto Nazionale di Fisica Nucleare, 
Italy; Ministry of Development of Turkey under Contract No. DPT2006K-120470; National Science and Technology fund; 
STFC (United Kingdom); The Royal Society, UK under Contracts Nos. DH140054, DH160214; The Swedish Research Council; 
U. S. Department of Energy under Contract No. DE-FG02-05ER41374.
\end{acknowledgments}

\nocite{*}

\bibliography{apssamp}

\end{document}